\documentclass{emulateapj}
\def\gtsim {\lower .1ex\hbox{\rlap{\raise .6ex\hbox{\hskip .3ex
        {\ifmmode{\scriptscriptstyle >}\else
                {$\scriptscriptstyle >$}\fi}}}
        \kern -.4ex{\ifmmode{\scriptscriptstyle \sim}\else
                {$\scriptscriptstyle\sim$}\fi}}}

\shorttitle{Redefining the Missing Satellites Problem}
\shortauthors{Strigari et~al.}

\begin{document}

\title{Redefining the Missing Satellites Problem}  

\author{Louis E. Strigari\altaffilmark{1}, 
 James S. Bullock\altaffilmark{1},
 Manoj Kaplinghat\altaffilmark{1},  
 Juerg Diemand \altaffilmark{2,5},  
 Michael Kuhlen\altaffilmark{3}, 
 Piero Madau\altaffilmark{2,4} 
 }
\altaffiltext{1}{Center for Cosmology,
Department of Physics \& Astronomy, University of California, Irvine, CA 92697}
\altaffiltext{2}{Department of Astronomy \& Astrophysics, University of 
California, Santa Cruz, CA 95064}
\altaffiltext{3}{School of Natural Sciences, Institute for Advanced Study, 
Einstein Drive, Princeton, NJ 08540}
\altaffiltext{4}{Max-Planck-Institut f\"ur Astrophysik, Karl-Schwarzschild-Str.
1, 85740 Garching, Germany.}
\altaffiltext{5}{Hubble Fellow}


\begin{abstract}

Numerical simulations of Milky-Way  size Cold Dark Matter (CDM)  halos
predict a  steeply  rising mass function   of {\em small}  dark matter
subhalos  and a  substructure count  that greatly  outnumbers the  observed
satellites of the Milky Way.  Several proposed explanations exist, but
detailed comparison between  theory  and observation  in terms  of the
maximum circular velocity ($V_{\rm max}$) of  the subhalos is hampered
by   the  fact that  $V_{\rm   max}$  for  satellite  halos is  poorly
constrained.   We present comprehensive mass  models for the well-known
Milky Way dwarf satellites, and   derive likelihood functions to  show
that their    masses   within $0.6$   kpc  ($M_{0.6}$)   are  strongly
constrained  by  the present  data. We show  that   the $M_{0.6}$ mass
function of luminous satellite halos  is flat between $\sim 10^7$  and
$10^{8}  M_{\odot}$.  We use the ``Via  Lactea'' N-body simulation to show
that the $M_{0.6}$ mass function of CDM subhalos is steeply rising
over this   range. We rule out the hypothesis that the 11 well-known satellites 
of the Milky Way are  hosted by the 11 most massive subhalos.  We show that  
models where the brightest satellites correspond to the  earliest  forming subhalos 
or the  most   massive accreted objects both reproduce the observed mass function.  
A similar analysis with the newly-discovered dwarf  satellites will further test
these   scenarios  and   provide  powerful  constraints   on  the  CDM
small-scale power spectrum and warm dark matter models.

\end{abstract}




\section{Introduction}

It  is now well-established that  numerical  simulations of Cold  Dark
Matter (CDM)  halos predict many orders of  magnitude more dark matter
subhalos around Milky Way-sized galaxies than observed dwarf satellite
galaxies \citep{klypin99,moore99,Diemand:2006ik}.  Within the  context
of the CDM  paradigm, there are well-motivated astrophysical solutions
to  this `Missing Satellites Problem' (MSP) that rely  on the idea that
galaxy formation is inefficient in the smallest dark matter halos
\citep{Bullock:2000wn,Benson:2001au,Somerville:2002,Stoehr:2002ht,Kravtsov:2004cm,Moore:2005jj,Gnedin:2006rt}. 
However,  from an observational perspective,  it has not been possible to 
distinguish between these solutions.

A detailed understanding of the MSP is limited by  our lack of a precise
ability  to characterize the dark matter  halos of satellite galaxies.
From  an observational perspective,  the primary constraints come from
the velocity dispersion of $\sim 200$ stars  in the population of dark
matter dominated dwarf spheroidal galaxies (dSphs)
\citep{Wilkinson:2004fz,Lokas:2004sw,Munoz:2005be,Munoz:2006hx,Westfalletal05,Walker:2005nt,Walker:2006qr,Sohn:2006et,Gilmore:2007fy}. 
However, the  observed extent of the stellar  populations in dSphs are
$\sim$ kpc, so these velocity dispersion measurements are only able to
probe properties    of the halos in this    limited radial regime.  

From   the   theoretical      perspective,  dissipationless  numerical
simulations typically characterize subhalo counts as a function of the
total bound  mass or maximum circular velocity,  $V_{\rm max}$.  While
robustly determined  in  simulations, global  quantities like  $V_{\rm
max}$  are difficult  to constrain observationally  because dark halos
can extend  well beyond the  stellar radius  of a  satellite.   Indeed
stellar kinematics alone provide only a  {\em lower limit} on the halo
$V_{\rm max}$ value (see below).   This is a fundamental limitation of
stellar kinematics that cannot be remedied by increasing the number of
stars     used    in     the     velocity      dispersion     analysis
\citep{Strigari:2007vn}.   Thus determining  $V_{\rm max}$ values  for
satellite    halos  requires   a     theoretical extrapolation.    Any
extrapolation of this kind is sensitive to the {\em predicted} density
structure   of subhalos, which   depends on  cosmology, power-spectrum
normalization, and the nature of dark matter \citep{Zentner:2003yd}.

Our  inability to determinate $V_{\rm  max}$ is the primary limitation
to test solutions to the MSP.  One particular solution states that the
masses of the dSphs have  been systematically underestimated, so  that
the $\sim 10$ brightest satellites  reside systematically in the $\sim
10$ most  massive  subhalos  \citep{Stoehr:2002ht,Hayashi:2002qv}.   A
byproduct of this solution is that  there must be  a sharp mass cutoff
at some {\em current}  subhalo mass, below  which galaxy formation  is
suppressed.  Other solutions, based on  reionization suppression, or a
characteristic halo  mass  scale {\em    prior to subhalo   accretion}
predict that  the suppression comes  in gradually with current subhalo
mass \citep{Bullock:2000wn,Kravtsov:2004cm,Moore:2005jj}.
 
In this paper, we  provide a   systematic investigation  of  the  masses  of the 
Milky Way satellites. We highlight that in  all cases the  {\em total} 
halo masses and maximum circular velocities are not  well-determined by the 
data. We instead use the fact that the  total cumulative mass within a  characteristic 
radius $\sim 0.6$ kpc is much  better  determined by the present data
\citep{Strigari:2007vn}.  We propose using  this mass, which we define
as  $M_{0.6}$,   as the  favored  means to  compare   to the dark halo
population in numerical simulations.  Unlike $V_{\rm max}$, 
$M_{0.6}$ is measured directly and requires no cosmology-dependent or 
dark-matter-model-dependent theoretical prior.

In the following sections,  we determine the $M_{0.6}$ mass
function  for  the  Milky   Way satellites,  and   compare  it to  the
corresponding mass function measured directly in the high-resolution 
``Via Lactea'' substructure simulation of \cite{Diemand:2006ik}. 
We rule out the possibility that there is a  one-to- one correspondence 
between the 11 most luminous satellites and the most massive subhalos in 
Via Lactea. We find that MSP  solutions based on reionization and/or 
characteristic halo mass scales prior to accretion are still viable.

\section{Milky Way Satellites} 
\label{sec:mwsatellites}
Approximately 20 satellite galaxies can be classified as residing  in
MW subhalos.   Of these, $\sim  9$ were discovered within the
last two years and have very low luminosities and surface brightnesses
\citep{Willman:2004kk,Willman:2005cd,Belokurov:2006hf,Belokurov:2006ph,
Zucker:2006bf,Zucker:2006he}. The  lack of precision  in these numbers
reflects the ambiguity in  the classification of the  newly-discovered
objects.  The nine `older' galaxies  classified as dSphs are supported
by  their velocity dispersion,     and exhibit no  rotational   motion
\citep{Mateo:1998wg}.  Two  satellite galaxies, the  Small  Megallanic
Cloud  (SMC)   and Large  Megallanic   Cloud (LMC),   are  most likely
supported by some combination of  rotation and dispersion. 
Stellar kinematics suggest that
the LMC and SMC are likely the most massive satellite systems of the
Milky Way.

We focus  on determining the  masses  of  the  nine most  well-studied
dSphs. The dark  matter masses of  the dSphs are determined from
the  line-of-sight   velocities of the  stars,   which trace the total
gravitational  potential.  We assume  a negligible contribution of the
stars to the gravitational potential, which we find to be an excellent
approximation.   The dSph with the    smallest mass-to-light ratio  is
Fornax, though even for this case we  find that the stars generally do
not affect    the    dynamics    of  the     system    \citep[see][and
below]{Lokas:2001mf,Wu:2007gt}.

Under the assumptions of equilibrium and spherical symmetry, the radial component of the stellar velocity dispersion, $\sigma_r$, is linked 
to the total gravitational potential of the system via the Jeans equation, 
\begin{equation}
\label{eq:jeans}
r \frac{d(\nu_{\star} \sigma_r^2)}{dr} =  - \nu_{\star} V_c^2
        - 2 \beta \nu_{\star} \sigma_r^2.
\end{equation} 
Here, $\nu_\star$ is the stellar density profile, $V_c^2 = GM(r)/r$ 
includes the total gravitating mass of the system,
 and the parameter $\beta (r) = 1 - \sigma_\theta^2/\sigma_r^2$ 
characterizes the difference between the radial ($\sigma_r$) and tangential 
($\sigma_\theta$) velocity dispersions. 
The observable velocity dispersion is constructed by integrating the 
three- dimensional stellar radial velocity dispersion profile along the 
line-of-sight, 
\begin{equation}
\sigma_{los}^2(R)  =   \frac{2}{I_\star(R)} \int_{R}^{\infty} \left ( 1 - \beta \frac{R^{2}}{r^2} \right )
\frac{\nu_{\star} \sigma_{r}^{2} r dr}{\sqrt{r^2-R^2}} \, , \label{eq:LOSdispersion}
\label{eq:sigmaLOS}
\end{equation}
where $R$ is the projected radius on the sky.
The surface density of stars in all dSphs are reasonably well-fit by a 
spherically-symmetric King profile \citep{King:1962wi}, 
\begin{equation}
I_\star(R) \propto  \left [ \left ( 1 + \frac{R^2}{r_{king}^2} \right )^{-1/2} -  
\left ( 1 + \frac{r_t^2}{r_{king}^2} \right )^{-1/2} \right ]^2, 
\label{eq:king}
\end{equation}
where $r_t$ and $r_{king}$ are fitting parameters denoted as the tidal
and core radii.~\footnote{Our results are insensitive to this particular 
parameterization of the light profile.}
The spherically symmetric stellar density can be obtained with an integral transformation 
of the surface density. From the form of Eq.~(\ref{eq:sigmaLOS}), the normalization in 
Eq.~(\ref{eq:king}) is irrelevant. 

Some dSphs show evidence for multiple, dynamically distinct stellar populations, with  each population  described  by its  own surface
density and velocity dispersion
\citep{Harbeck:2001bu,Tolstoy:2004vu,Westfalletal05,McConnachie:2006nb,Ibata:2006xr}.
In a dSph with $i = 1 ... N_p$ populations of stars, standard observational
methods will sample a density-weighted average of the populations:
\begin{eqnarray}
\nu_\star & = & \sum_{i} \nu_i \, \\
\sigma_r^2  & = & \frac{1}{\nu_\star}\sum_{i} \nu_i \sigma_{r,\imath}^2 \, ,
\end{eqnarray}
where $\nu_i$ and $\sigma_i$ are the density profile and radial stellar velocity dispersion of stellar component $i$.  
In principle, each component has its own stellar velocity anisotropy profile,
$\beta_i(r)$.  In this case,  Equation~(\ref{eq:jeans}) is valid for $\nu_\star$ and $\sigma_r$ as long as an effective 
velocity anisotropy is defined as
\begin{eqnarray}
\beta(r)     & = & \frac{1}{\nu_\star \sigma_r^2} \sum_{i} \beta_i \nu_i \sigma_i^2. 
\end{eqnarray}
From these definitions, we also have $\sum_\imath I_{\star,\imath} \sigma_{los,\imath}^2 = I_\star \sigma_{los}^2$. 

The conclusion  we draw  from this  argument  is that the presence  of
multiple populations will  not   affect the  inferred mass
structure  of the   system,
provided that the  velocity anisotropy is modeled as a free
function of radius. Since the functional form  of the stellar velocity
anisotropy is unknown, we  allow for a  general, three parameter model
of the velocity anisotropy profile,
\begin{equation}
\beta (r) = \beta_\infty \frac{r^2}{r_\beta^2 + r^2} + \beta_0. 
\label{eq:betaprofile}
\end{equation}
A profile of this form 
allows for the possibility for $\beta(r)$ to change from radial to tangential orbits 
within the halo, 
and a constant velocity anisotropy is recovered in the limit 
$\beta_\infty \rightarrow 0$, and $\beta_0 \rightarrow$ const. 

In Equations~(\ref{eq:jeans}) and (\ref{eq:sigmaLOS}), the radial stellar velocity dispersion, $\sigma_r$, 
depends on the total mass distribution, and thus the parameters
describing the dark matter density profile.
Dissipation-less N-body simulations show that the density profiles of CDM halos 
can be characterized as 
\begin{equation}
\rho (\tilde{r}) = \frac{\rho_s}{\tilde{r}^{\gamma} (1+\tilde{r})^{\delta}}; \hspace{0.6cm} 
\tilde{r} = r / r_s, 
\label{eq:nfw}
\end{equation}
where $r_s$ and $\rho_s$ set a radial scale and density normalization
and  $\gamma$   and $\delta$ parameterize  the  inner and
outer slopes of the  distribution.  For dark matter halos unaffected by tidal interactions, 
the most recent high-resolution simulations  find $\delta  + \gamma \approx  3$ works well for the 
outer slope, while $0.7    \lesssim  \gamma \lesssim      1.2$ works well down to 
$\sim 0.1 \%$ of halo virial radii \citep{Navarro:2003ew,Diemand:2005wv}.
This interior slope is not altered as a subhalo loses mass from tidal
interactions   with  the MW potential \citep{Kazantzidis:2005su}. The outer slope, $\delta$, 
depends more sensitively on the tidal interactions in the halo. The majority of the 
stripped material will be from the outer  parts of halos, and thus  
$\delta$ of subhalo density  profiles will  become  steeper than  those  of  field  
halos. Subhalos are characterized by outer slopes in the range 
$2    \lesssim  \delta \lesssim      3$. 

Given the uncertainty in the $\beta(r)$ and $\rho(r)$ profiles, we are
left  with  nine  shape  parameters  that  must  be   constrained  via
line-of-sight   velocity dispersion  measurements:   $\rho_s$,  $r_s$,
$\beta_0$,  $\beta_\infty$, $r_\beta$, $\gamma$, $\delta$, $r_{king}$,
and $r_t$.  While the problem as posed may  seem impossible, there are
a   number of physical  parameters,    which  are degenerate   between
different profile  shapes, that are   well constrained.  Specifically,
the stellar populations constrain  $V_c(r)$ within a radius comparable
to the stellar radius  $r_t \sim$ kpc.   As a result,  quantities like
the local  density and integrated mass  within  the stellar radius are
determined with   high precision   \citep{Strigari:2006rd},      while
quantities  that depend on the mass   distribution at larger radii are
virtually unconstrained by the data.

It is useful to determine the value of the radius where the
constraints are maximized.
The location of this characteristic 
radius is determined by the form of the 
integral in Eq.~(\ref{eq:sigmaLOS}). We can gain some insight 
using the example of a power-law stellar distribution $\nu_\star(r)$,
power-law dark matter density profile $\rho \propto r^{-\gamma_\star}$, 
and constant velocity anisotropy. The line-of-sight velocity dispersion depends on the 
three-dimensional stellar velocity dispersion, which can be written as
\begin{equation}
\sigma_r^2(r) = \nu_\star^{-1}  r^{-2 \beta} \int_r^{r_t} G
\nu_\star(r) M(r) r^{2 \beta - 2} {\rm d}r \propto r^{2 -
  \gamma_\star} 
\label{eq:3Ddispersion}  
\end{equation}  
From the shape of the King profile, the majority of stars reside at projected radii 
$r_{\rm king} \lesssim R \lesssim r_{\rm t}$, 
where the stellar distribution is falling rapidly $\nu_\star \sim r^{-3.5}$.  In this
case, for  $\beta = 0$, the LOS component scales as $\sigma_{los}^2(R)
\propto \int_R^{r_t} r^{-0.5 - \gamma_\star} (r^2 - R^2)^{-1/2} {\rm
  d}r$ and is dominated by the mass and density profile at the
smallest relevant radii, $r \sim r_{\rm king}$. For $R \lesssim r_{\rm
  king}$, $\nu_\star \propto r^{-1}$ and $\sigma_{los}^2$ is similarly
dominated by $r \sim r_{\rm   king}$ contributions.  
We note  that although  the  scaling arguments above
hold   for constant velocity  anisotropies, they   can be extended  by
considering  anisotropies that vary significantly  in radius. They are
also independent of the precise form of $I_\star$, provided there is a
scale  similar  to  $r_{king}$. 

In   \cite{Strigari:2007vn}, it was shown that
typical velocity dispersion profiles best constrain the mass (and density)
within a characteristic radius $\simeq 2 \, r_{king}$.   
 For  example,  the  total  mass  within  $2
r_{king}$ is constrained  to within $\sim 20\%$  for dSphs  with $\sim
200$ line-of-sight  velocities.  Note  that when  deriving constraints
using  only  the innermost stellar   velocity dispersion and fixing the
anisotropy   to  $\beta  =  0$,  the characteristic  radius  for  best
constraints decreases to $\sim 0.5 r_{king}$ \citep[e.g.][]{Penarrubia:2007zz}.

As listed in Table~\ref{tab:parameterstable}, the 
Milky Way dSphs are observed to  have  variety of $r_{king}$  values, 
but $r_{king} \sim 0.3$ kpc is typical. The values of $r_{king}$ and $r_t$
are taken from \cite{Mateo:1998wg}. 
In order to facilitate comparison with simulated subhalos, we chose
a single characteristic  radius of 0.6 kpc for all the dwarfs, 
and  we represent the mass within this
radius as $M_{0.6} = M(< 0.6 \, {\rm kpc})$. The relative
errors on the derived masses are unaffected for small variations  in the 
characteristic radius in the range $\sim 1.5-2.5 \, r_{king}$. 
Deviations from a true King
profile at large radius (near $r_t$) do not affect these arguments,  as long as there is
a  characteristic  scale   similar   to  $r_{king}$  in   the  surface
density. The only dSph significantly affected by the choice of 0.6 kpc
as   the characteristic radius  is  Leo  II,   which  has $r_t =  0.5$
kpc. Since the characteristic radius is greater than twice  $r_{king}$, the 
constraints on its mass will be weakest of  all  galaxies (with the exception of
Sagittarius, as discussed below).

\section{Dark Matter Halo Masses at the Characteristic Radius}
\label{sec:constraints} 

We use the following data sets: 
\cite{Wilkinson:2004fz,Munoz:2005be,Munoz:2006hx,Westfalletal05,Walker:2005nt,Walker:2006qr,Sohn:2006et};  
Siegal et al. in preparation. 
These velocity dispersions are determined from the line-of-sight velocities of $\sim 200$ stars
in each galaxy, although observations in the 
coming years will likely increase this number by a factor $\sim 5-10$. 
From the data, we calculate the $\chi^2$, defined in our case as 
\begin{equation} 
\chi^2 =  \sum_{\imath=1}^n \frac{(\sigma_{obs,\imath} - \sigma_{th,\imath})^2}{\epsilon_\imath^2}. 
\label{eq:chisquared}
\end{equation}
Here $\sigma_{obs}^2$ is the observed velocity dispersion in each bin, $ \sigma_{th}^2$
is the theoretical value, obtained from Eq.~(\ref{eq:LOSdispersion}), and $\epsilon_\imath^2$
are errors as determined from the observations. 

It is easy to see that, when fitting to a single data set of $\sim 200$ stars, parameter 
degeneracies will be significant. However, from the discussion in 
section~\ref{sec:mwsatellites}, $M_{0.6}$ is well-determined by the LOS data. 
To determine how well $M_{0.6}$ is constrained, we construct likelihood functions for each galaxy.
When thought of as a function of the theoretical parameters, the likelihood 
function, ${\cal L}$, is defined as the probability that a data set is acquired given a
set of theoretical parameters. 
In our case ${\cal L}$ is a function of the parameters $\gamma$, $\delta$, $r_s$, 
$\rho_s$, and $\beta_0$, $\beta_\infty$, $r_\beta$, and is defined as
${\cal L} = e^{-\chi^2/2}$.  
In writing this likelihood function, we assume that the errors on the measured 
velocity dispersions are Gaussian, which we find to be an excellent approximation to the 
errors for a dSph system \citep{Strigari:2007vn}. We marginalize over the parameters
$\gamma$, $\delta$, $r_s$, $\rho_s$, and $\beta_0$, $\beta_\infty$, $r_\beta$ at fixed
$M_{0.6}$, and the optimal values for $M_{0.6}$ are determined by the maximum of ${\cal L}$ .

We determine  ${\cal L}$ for all  nine dSphs  with velocity dispersion
measurements.  For all galaxies we use the full published velocity 
dispersion profiles. 
The only galaxy that does not have a published velocity dispersion profile
is Sagittarius, and for this galaxy we use the central velocity dispersion 
from \citet{Mateo:1998wg}. 
The mass modeling of Sagittarius is further complicated by the  fact that it is
experiencing tidal interactions  with the  MW
\citep{Ibata:1996dv,Majewski:2003ux},
so a mass   estimate from   the  Jeans equation  is not    necessarily
reliable. We caution that in this case the mass we determine is likely
only an approximation to the total mass of the system.

We determine the likelihoods by marginalizing over the following ranges of
the velocity anisotropy, inner and outer slopes: $-10 < \beta_0 < 1$, $-10 < \beta_\infty < 1$,
$0.1 < r_\beta < 10$ kpc, $0.7 < \gamma < 1.2$, and $2 < \delta < 3$. 
As discussed above, these ranges for the asymptotic inner and outer slopes are appropriate 
because we are considering CDM halos. It is important to emphasize that
these ranges are theoretically motivated and that
observations alone do not demand such restrictive choices.
It is possible to fit all of the dSphs at present with 
a constant density cores with scale-lengths $\sim 100$ pc 
\citep{Strigari:2006ue,Gilmore:2007fy}, although the data by no
means demand such a situation. Though we consider inner 
and outer slopes in the ranges quoted above, 
our results are not strongly affected if
we widen these intervals. 
For example, we find that if we allow the inner slope to vary down to 
$\gamma = 0$, the widths of the likelihoods are only changed by 
$\sim 10\%$. This reflects the fact that there is a negligible
degeneracy between $M_{0.6}$ and the inner and outer slopes. 

\begin{deluxetable*}{l|lllllllc}
\tabletypesize{\scriptsize}
\tablewidth{0pt}
\tablecaption{Parameters Describing Milky Way Satellites. 
\label{tab:parameterstable} }
\tablehead{
\colhead{Galaxy}  & \colhead{$r_{king}$} & \colhead{$r_t$} & 
\colhead{L$_V$}  & \colhead{Mass $< 0.6$ kpc} 
 & \colhead{Mass $< r_t$} 
& \colhead{$M(<r_t)/L$} & 
\colhead{$V_{\rm max} \, [{\rm km} \, {\rm s}^{-1}]$} & 
\colhead{$V_{\rm max} \, [{\rm km} \, {\rm s}^{-1}]$}   \\
&[kpc]  & [kpc]  & [10$^6$ L$_\odot$]& $[10^7 \, {\rm M}_\odot$]
& $[10^7 \, {\rm M}_\odot$]
& [${\rm M}_\odot/L_\odot$]&  (w/o prior) & (with theory prior) } 

\startdata
Draco & 0.18 & 0.93 & $0.26$ & $4.9^{+1.4}_{-1.3}$&$14^{+7.0}_{-4.2}$ 
&530& $> 22$ & $28_{-9}^{+21}$ \\
Ursa Minor & 0.30& 1.50 & $0.29$& $5.3^{+1.3}_{-1.3}$ & $23^{+16}_{-11}$
&790& $>21$ & $26_{-6}^{+12}$ \\
Leo I & 0.20  & 0.80 & 4.79 & $4.3^{+1.6}_{-1.6}$ & $8.5^{+4.5}_{-2.8}$
&106 & $>14$ & $19_{-5}^{+13}$ \\
Fornax & 0.39 & 2.70 & $15.5$ & $4.3^{+2.7}_{-1.1}$& $44^{+31}_{-29}$
&28 &$>20$ & $25_{-5}^{+5}$ \\
Leo II & 0.19 & 0.52 & 0.58  & $2.1^{+1.6}_{-1.1}$&$2.1^{+1.6}_{-1.1}$ &128& 
$ > 17$ & $9^{+3}_{-1}$ \\
Carina & 0.26 & 0.85  & $0.43$ & $3.4^{+0.7}_{-1.0}$ &$6.7^{+2.3}_{-2.5}$  
& 82& $>13$ &  $15_{-3}^{+5}$  \\
Sculptor & 0.28& 1.63 & $2.15$& $2.7^{+0.4}_{-0.4}$& $15^{+0.7}_{-1.5}$
&68 & $ >20$ & $14_{-2}^{+2}$ \\
Sextans & 0.40 & 4.01 &  $0.50$  & $0.9^{+0.4}_{-0.3}$& $13^{+11}_{-5.8}$
&260& $ >8$ & $9^{+1}_{-1}$  \\
Sagittarius & 0.3 & 4.0 &18.1  & $20^{+10}_{-20}$ & $>20$ &$>11$& $> 19$ & --- \\
\enddata
\tablecomments{Determination of the mass within 0.6 kpc and the maximum circular 
velocity for the dark matter halos of the dSphs. The errors are determined as the location
where the likelihood function falls off by $90\%$ from its peak value. For Sagittarius, no
reliable estimate of $V_{\rm max}$ with the CDM prior could be determined.
The CDM prior is determined using the concordance cosmology with $\sigma_8 = 0.74$, 
$n = 0.95$ (see text for details). 
}
\end{deluxetable*}

\begin{figure}
\plotone{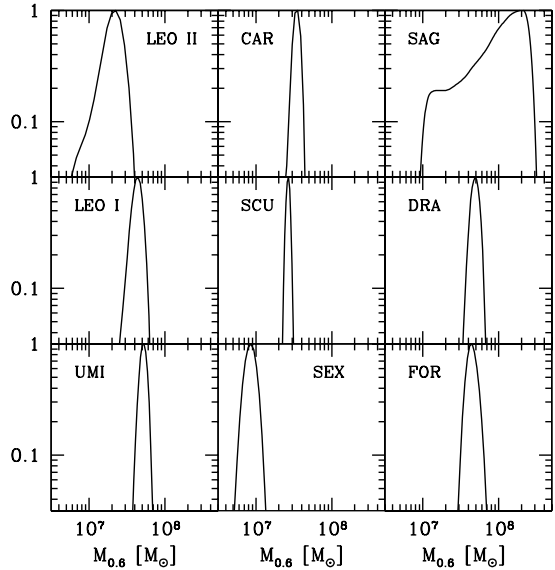}
\caption{\label{fig:m0.6kpc} 
The likelihood functions for the mass within 0.6 kpc for the nine dSphs, normalized to unity at the peak. 
 }
\end{figure} 

We are left to determine the regions of $\rho_s$   and $r_s$ parameter
space to marginalize over. In  all  dSphs, there  is  a degeneracy in this
parameter space, telling us that it is not possible to derive an upper
limit  on      this  pair of    parameters    from    the  data  alone
\citep{Strigari:2006rd}. While  this degeneracy is not  important when
determining constraints on  $M_{0.6}$, it is  the primary  obstacle in
determining $V_{\rm   max}$. From the  fits we  present below, we find
that the lowest  $r_s$ value that provides an acceptable
fit  is $\sim 0.1$ kpc,  and  we use this  as the  lower limit in all
cases.  In  our fiducial mass  models,  we conservatively restrict the
maximum value of $r_s$ using the known  distance to each dSph. In this
case, we use  $0.1 \, {\rm kpc} < r_s < D/2$,   where $D$ is   the distance to  the
dSph. 

\begin{figure}
\plotone{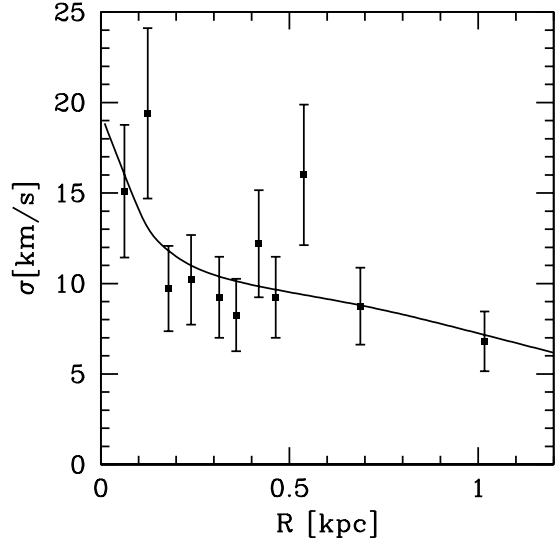}
\caption{\label{fig:sigma} 
The velocity dispersion for Ursa Minor as a function of radial distance, along with
the model that maximizes the likelihood function.  
 }
\end{figure}

\begin{figure}
\plotone{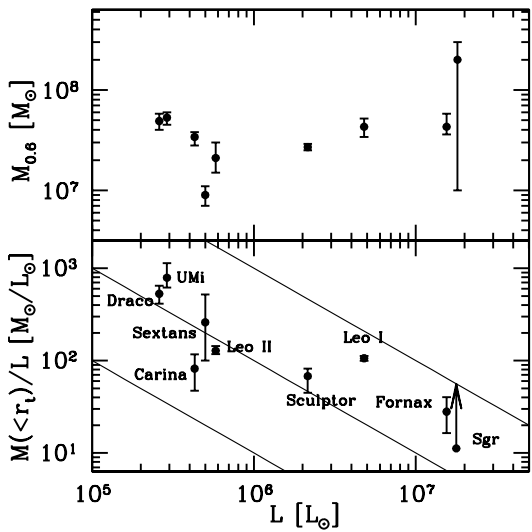}
\caption{\label{fig:sum} 
The mass within 0.6 kpc (upper) and the mass-to-light ratios within the King tidal radius (lower) 
for the Milky Way dSphs as a function of dwarf luminosity. 
The error-bars here are defined as the locations where the likelihoods fall to $40\%$ of the peak 
values (corresponding to $\sim 1 \sigma$ errors). The lines denote, from top to bottom, constant
values of mass of $10^7, 10^8, 10^9 \, M_\odot$. 
 }
\end{figure}

In Figure~\ref{fig:m0.6kpc} we show the $M_{0.6}$ likelihood functions for 
all of  the dSphs.
As is shown,  we obtain strong constraints on $M_{0.6}$
in all  cases except Sagittarius, for which we use only a central
velocity dispersion. Table~\ref{tab:parameterstable} summarizes the best 
fitting $M_{0.6}$ values for each dwarf. The quoted errors correspond to
the points where the likelihood falls to $10\%$ of its peak value. The upper panel 
of Figure \ref{fig:sum} shows $M_{0.6}$ values for each dwarf as a function of 
luminosity. In Figure~\ref{fig:sigma} we show an example of the velocity dispersion
data as a function of radial distance for Ursa Minor, along with the model that
maximizes the likelihood function. For all galaxies, we find $\chi^2$ per degree 
of freedom values $\lesssim 1$. 

The maximum likelihood method also allows us to constrain the mass at 
other radii spanned by the stellar distribution.  The sixth column
of Table 1 provides  the integrated mass within each dwarf's King tidal radius. 
This radius roughly corresponds to the largest radius where a reasonable mass 
constraint is possible. As expected, the mass within $r_t$ is not as well determined 
as the mass within $2 \, r_{king}$. From these masses we are able to determine the 
mass-to-light ratios within $r_t$, which we present in the seventh column of 
Table~\ref{tab:parameterstable}. In the  bottom panel  of Figure~\ref{fig:sum}, we   
show mass-to-light ratios within  $r_t$ as a function  of  dwarf luminosity. We see 
the standard result that the observable mass-to-light ratio increases with
decreasing luminosity \citep{Mateo:1998wg}. Note, however, that  our results
are  inconsistent with the  idea that all of the  dwarfs have the same
integrated mass within their stellar extent. We note that for Sagittarius, 
we can only obtain a lower limit on the total mass-to-light ratio. 

The last two columns in Table 1 list constraints on $V_{\rm max}$ for
the dSphs. Column 8 shows results for an analysis with limits on 
 $r_s$ as described above.  In this case, the integrated mass within the 
 stellar radius is constrained by the velocity dispersion data, but the halo 
 rotation velocity curve, $V_c(r)$, can continue to rise as $r$ increases 
 beyond the stellar radius in an unconstrained manner.
The result is that the velocity dispersion data alone provide only 
a lower limit on $V_{\rm max}$.

Stronger constraints on $V_{\rm max}$ can be obtained if we 
limit the range of $r_s$  by imposing a cosmology-dependent  
prior on the dark matter mass profile. CDM simulations 
have shown that there is a correlation between 
  $V_{\rm max}$ and $r_{\rm max}$ for halos,
where $r_{\rm max}$ is the radius where the circular velocity peaks.
Because subhalo
densities will  depend on the collapse  time and  orbital evolution of
each system,  the   precise $V_{\rm max}$-$r_{\rm max}$    relation is
sensitive to cosmology (e.g. $\sigma_8$)  and the formation history of
the host halo itself 
\citep[e.g.][]{Zentner:2003yd,Power03,Kazantzidis:2005su,BJ:05,BJ:06}.
 When  converted to the relevant  halo parameters, the imposed $V_{\rm
 max}$-$r_{\rm max}$ relation can  be seen as  a theoretical prior  on
 CDM halos, restricting the parameter space we need to integrate over.
 In  order  to illustrate   the technique, we  adopt   $\log_{10} ( r_{\rm  max} ) =
 1.35(\log_{10} ( V_{\rm max}/{\rm km}\,{\rm s}^{-1} ) -1) -0.196$ kpc with a  
 scatter of $0.2$ in $\log_{10}$, as measured from  simulated subhalos within the  
 Via Lactea host halo \citep{Diemand:2007qr}. This simulation is for a LCDM
cosmology with  $\sigma_8 = 0.74$ and $n=0.95$. The scatter in the subhalo mass
 function increases at the very high mass end, which reflects the fact that these most
 massive subhalos are those that are accreted most recently 
 \citep{Zentner:2003yd,vandenBosch:2004zs}. However, as we show below our results 
 are not strongly dependent on the large scatter at the high mass end.  
 
Column 9 in Table 1 shows the allowed subhalo
$V_{\rm max}$ values for the assumed prior.
Note that in most cases, this prior degrades the quality of the fit, and 
the likelihood functions peak at a lower overall value. 
The magnitude of this effect is not large except for the cases of 
Leo II and Sagittarius.  For Leo II, the peak likelihood with the
prior occurs at a value that is below the $10 \%$ likelihood
for the case without a prior on $r_s$ (i.e. the data seem to
prefer a puffier subhalo than would be expected in CDM).
For Sagittarius, we are unable to obtain a reasonable fit within a
subhalo that is typical of those expected.  This is 
not too surprising.  Sagittarius is being tidally disrupted 
and its dark matter halo is likely atypical. 

 We emphasize that the  $V_{\rm max}$ determinations 
listed in Column 9 are driven by {\em theoretical} 
assumptions, and can only be fairly 
compared to predictions for
this specific cosmology (LCDM, $\sigma_8 = 0.74$).  
The $M_{0.6}$ values in Column 5 are applicable
for any theoretical model, including non-CDM models, or CDM 
models with any normalization or power spectrum shape.

\section{Comparison to Numerical Simulations}

The   recently-completed  Via Lactea  run  is   the highest-resolution
simulation  of galactic substructure  to date, containing  an order of
magnitude more particles than its predecessors \citep{Diemand:2006ik}.
As  mentioned above,  Via  Lactea   assumes  a  LCDM cosmology   with
$\sigma_8 =  0.74$ and $n= 0.95$.  For  a detailed description  of the
simulation,  see  \cite{Diemand:2006ik}.   For our purposes,  the most
important aspect of Via  Lactea is its ability  to resolve the mass of
subhalos on length  scales of the  characteristic radius 0.6  kpc.  In
Via Lactea,   the force    resolution   is 90  pc  and   the  smallest
well-resolved length scale is 300 pc, so  that the mass within 0.6 kpc
is well-resolved in nearly  all subhalos.  
Due to the choice of time steps we expect the simulation to underestimate local
densities in the densest regions (by about $10 \%$ at densities of $9 \times
10^7 \, {\rm M}_\odot/{\rm kpc}^3$). There
is only one subhalo with a higher local density  than this at 0.6 kpc.
For this subhalo, $\rho(r=0.6 \, {\rm kpc})  = 1.4 \times 10^8 \, {\rm
M}_\odot/{\rm kpc}^3$, so its local density might be underestimated by
up to $10\%$, and the errors in the enclosed mass might be $\sim 20\%$
\citep{Diemand:2005wv}. For  all other subhalos  the  densities at 0.6
kpc  are well  below the affected  densities,  and  the enclosed  mass
should not be  affected by more than $10\%$   by the finite  numerical
resolution.

We define subhalos in Via Lactea to be the self-bound halos that
lie within the radius ${\rm R}_{200} = 389$ kpc, where ${\rm R}_{200}$ is defined to 
enclose an average density 200 times the mean matter density.  We note that in 
comparing to the observed MW dwarf population, we could have conservatively chosen 
subhalos that are restricted to lie within the same radius as the most distant MW dSph 
(250 kpc).  We find that this choice has a  negligible effect on our conclusions --
it reduces the count of small halos by $\sim 10\%$.    

\begin{figure}
\plotone{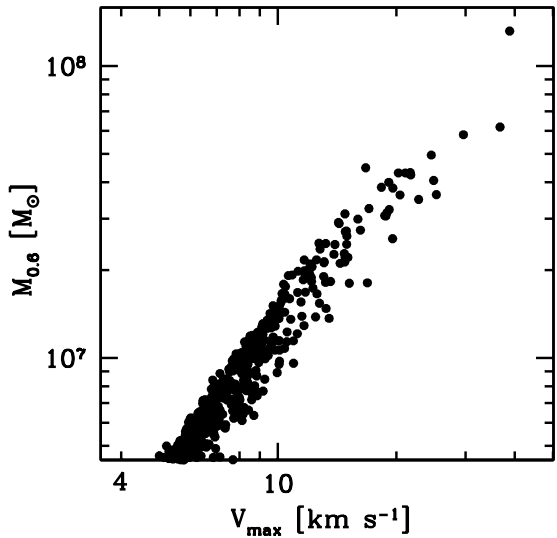}
\caption{\label{fig:m06vsvmax} 
The mass within 0.6 kpc versus the maximum circular velocity for the mass ranges of Via
Lactea subhalos corresponding to the population of satellites we study. 
 }
\end{figure}

In Figure~\ref{fig:m06vsvmax}, we show how
$M_{0.6}$ relates to the more familiar quantity 
$V_{\rm max}$ in Via Lactea subhalos.
We note that the relationship between subhalo $M_{0.6}$ and $V_{\rm max}$
will be sensitive to the power spectrum shape and normalization,
as well as the nature of dark matter~\citep{Bullock:1999he,Zentner:2003yd}.
The relationship shown is only valid for the Via Lactea cosmology, but
serves as a useful reference for this comparison.

Given likelihood functions for the dSph $M_{0.6}$ values,
we are now in position 
to determine the $M_{0.6}$ mass function for Milky Way (MW)
 satellites and compare this to the 
corresponding mass function in Via Lactea.
For both the observations and the simulation, we count the number of systems
in four mass bins from $4 \times 10^6 < M_{0.6} <  4 \times 10^8 \, {\rm M}_\odot$. 
This mass range is chosen to span the $M_{0.6}$ values allowed by
the likelihood functions for the MW satellites. 
We assume that the two non-dSph satellites, the LMC and SMC, belong in the 
highest mass bin, corresponding to 
$M_{0.6} > 10^8 \, {\rm M}_\odot$ \citep{Harris:2006cr,vdMarel:02}. 

\begin{figure}
\plotone{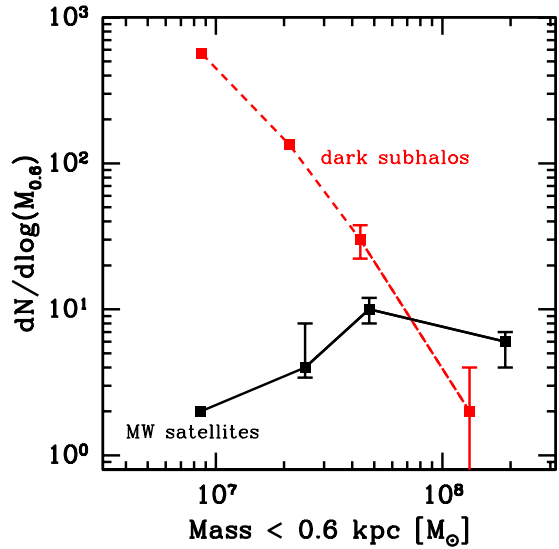}
\caption{\label{fig:nm1} 
The $M_{0.6}$ mass function of Milky Way satellites and dark subhalos in 
the Via Lactea simulation. The red (short-dashed) curve is the total subhalo mass function
from the simulation. The black (solid) curve is the median of the observed 
satellite mass function. The 
error-bars on the observed mass function represent the upper and lower
limits on the number of configurations that occur with a probability of $> 10^{-3}$.
 }
\end{figure}

In  Figure~\ref{fig:nm1} we show  resulting   mass functions for  MW
satellites (solid) and for  Via Lactea subhalos (dashed, with Poisson error-bars). 
For  the  MW satellites, the
central values  correspond to the  median number of  galaxies per bin,
which  are  obtained  from   the maximum   values   of the  respective
likelihood functions. The error-bars on the satellite points
are set by the upper and
lower configurations   that occur with   a probability of  $> 10^{-3}$
after   drawing  1000  realizations  from  the   respective likelihood
functions.   As seen in Figure~\ref{fig:nm1}, the predicted dark subhalo
mass function rises as $\sim M_{0.6}^{-2}$ while the visible MW satellite
mass function is relatively flat.  The lowest mass bin
($M_{0.6} \sim 9 \times 10^6 M_{\odot}$) 
always contains 1 visible galaxy (Sextans).  The second-to-lowest mass bin
($M_{0.6} \sim 2.5 \times 10^7 M_{\odot}$) contains between 2 and 4 satellites
(Carina, Sculptor, and Leo II). The  fact that these  two  lowest bins are
not consistent with zero galaxies has important implications for the
\cite{Stoehr:2002ht} solution to the MSP: specifically, it implies that 
the 11 well-known MW satellites do not reside in subhalos that resemble the
11 most massive subhalos in Via Lactea. 

To further emphasize this point, we see from Figure~\ref{fig:nm1} that the mass of the 
11th most massive subhalo in Via Lactea is $4 \times 10^7 \, {\rm M}_\odot$. From
the likelihood functions in Figure~\ref{fig:m0.6kpc}, Sextans, Carina, Leo II, and Sculptor
must have values of $M_{0.6}$ less than $4 \times 10^7 \, {\rm M}_\odot$ at $99 \%$ c.l., 
implying a negligible probability that all of these dSphs reside in halos
with $M_{0.6} > 4 \times 10^7 \,  {\rm M}_\odot$. 

\begin{figure}
\plotone{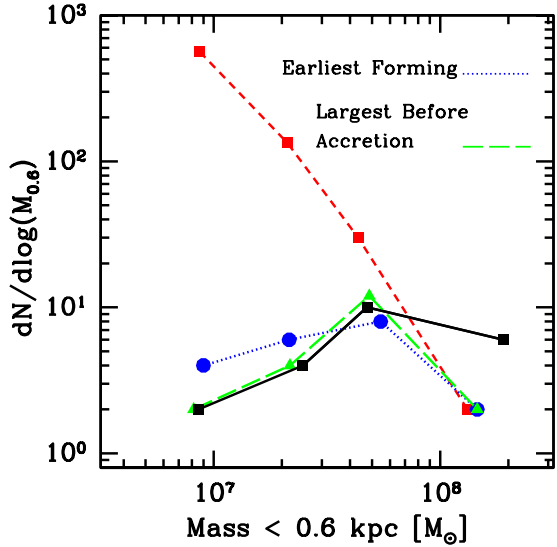}
\caption{\label{fig:nm2} 
The solid and dashed curves show the MW satellites and
dark subhalos in Via Lactea, respectively.   These lines are reproduced from
Figure~\ref{fig:nm1}, with error-bars suppressed for clarity. The blue (dotted) 
curve represents the ten earliest forming halos in Via Lactea, and the green 
(long-dashed) curve represents the 10 most massive halos before accretion 
into the Milky Way halo. 
 }
\end{figure}

Using the $M_{0.6}$ mass function of MW satellites, we can test other CDM-based 
solutions to the MSP. 
Two models of interest are based on 
reionization  suppression \citep{Bullock:2000wn,Moore:2005jj} and  on there being a
characteristic  halo mass   scale  prior   to subhalo  accretion \citep{Diemand:2006ik}.
 To roughly represent these models, we 
focus on two subsamples of Via Lactea subhalos: the earliest forming (EF) halos, 
and the largest mass halos before they were accreted (LBA) into the host halo. 
As described in \cite{Diemand:2006ik}, the LBA sample is defined to be the 10 subhalos 
that had the highest $V_{\rm max}$ value throughout their entire history. These systems all had 
$V_{\rm max} > 37.3 \, {\rm km} \, {\rm  s}^{-1}$ at some point in their history. The EF sample consists 
of the 10 subhalos with $V_{\rm max} > 16.2 \, {\rm km} \, {\rm  s}^{-1}$ (the limit of atomic cooling) 
at $z=9.6$.  The \cite{Kravtsov:2004cm} model would correspond to a selection intermediate between 
EF and LBA. In Figure~\ref{fig:nm2} we show the observed mass 
function of MW satellites (solid, squares) along with
the EF (dotted, triangles) and LBA (long-dashed, circles) samples. 
We conclude that both of these models 
are in agreement with the MW satellite mass function. Future observations and 
quantification of the masses of the newly-discovered MW satellites will enable us 
to do precision tests of the viable MSP solutions. Additionally, once the capability to
do numerical simulations of substructure in warm dark matter models becomes a 
reality, the $M_{0.6}$ mass function will provide an invaluable tool to place 
constraints on WDM models.  

\section{Summary and Discussion} 

We have provided comprehensive dark  matter mass constraints for the 9
well-studied dSph satellite galaxies of the Milky Way and investigated
CDM-based  solutions for  the missing satellite  problem  in  light of
these  constraints.    While  subhalo $V_{\rm  max}$   values  are the
traditional  means     by   which theoretical     predictions quantify
substructure counts, this is not  the most direct  way to confront the
observational   constraints.  Specifically, $V_{\rm  max}$  is  poorly
constrained by stellar  velocity dispersion measurements, and can only
be estimated     by   adopting   cosmology-dependent,      theoretical
extrapolations.     We  argue the    comparison   between  theory  and
observation is  best  made using the integrated  mass   within a fixed
physical   radius  comparable to  the  stellar   extent of  the  known
satellites, $\sim   0.6$  kpc.   This  approach is  motivated  by
\cite{Strigari:2007vn} who  showed that the   mass within two  stellar
King radii is best  constrained  by typical velocity dispersion  data.

Using $M_{0.6}$  to represent the dark  matter mass within a radius of
0.6 kpc,  we computed $M_{0.6}$  likelihood functions for the MW dSphs
based  on published velocity dispersion  data. Our models allow for a
wide range of underlying dark matter halo profile shapes and stellar
velocity dispersion profiles.  With this broad allowance, we showed that
the $M_{0.6}$ for most dwarf satellites is constrained to within
$\sim 30 \%$.

We derived the $M_{0.6}$  mass function of MW satellites (with
error  bars)  and compared  it    to the same  mass  function  computed
directly from the Via Lactea substructure simulation.  While the 
observed $M_{0.6}$ mass function of luminous satellites is relatively flat,
the comparable CDM subhalo mass function rises as $\sim M_{0.6}^{-2}$.  
We rule out the hypothesis that all of the well-known
 Milky Way satellites strictly inhabit the most 
massive CDM subhalos.  If luminosity does track {\em current} subhalo mass,
this would only be possible if the subhalo population of the 
Milky Way were drastically different than that predicted in CDM.
However, we show that other plausible CDM solutions are consistent with
the observed mass function. Specifically, the earliest forming subhalos 
have a flat $M_{0.6}$ mass function that is consistent with the
satellite subhalo mass function.  This would be expected 
if the population of bright dwarf spheroidals corresponds to the residual halo population 
that accreted a significant mount of gas 
before the epoch of reionization \citep{Bullock:2000wn}. 
We also
tested the hypothesis that the present dwarf spheroidal population corresponds to the subhalos 
that were the most massive before they fell into the MW halo \citep{Kravtsov:2004cm}.
This hypothesis is also  consistent with the current data. 

In deriving the $M_{0.6}$ mass function for this paper we have set aside the
issue of the most-
recently discovered  population of MW dwarfs. We aim to return to this
issue in later work, but it is worth speculating on the expected
impact that these systems would have on our conclusions.
  If we had included the new systems,  making  $\sim 20$  satellites in
all, would it be possible to place these systems in the $\sim 20$ most
massive subhalos in Via  Lactea?  Given the  probable mass ranges  for
the new  dwarfs, we find that  this is unlikely.   We can  get a rough
estimate of their masses from their observed luminosities. We start by
considering the mass-to-light ratios of the known dSph population from
figure~\ref{fig:sum} and  from \cite{Mateo:1998wg}.  If we assume that
the other dwarfs have similar $M/L$ range, we  can assign a mass range
for each of them. In all cases, the  new MW dwarfs are approximately 1
to 2  orders of magnitude  smaller  in luminosity than  the well-known
dSph population.  Using the central  points for  the  known dSphs,  we
obtain $M_{0.6}/L$ spanning the range from $3-230$. Considering the width of
the likelihoods, we can allow a  slightly larger range, $2-350$. If we
place the new  dwarfs in this latter range,  the uncertainty in  their
masses is $(2-350) L  M_\odot/L_\odot$. Even with this generous  range
we expect most of the new dwarfs  have $M_{0.6} \lesssim 10^7 \, {\rm M}_\odot$. 
\footnote{These estimates are in rough agreement with recent determinations
from  stellar velocity dispersion measurements in  the  new dwarfs, as
presented  by N.  Martin  and  J. Simon at  the  3rd  Irvine Cosmology
Workshop,              March               22-24,                2007,
http://www.physics.uci.edu/Astrophysical-Probes/}   

The discovery of more members of  the MW group, and the precise 
determination  of the $M_{0.6}$ mass function, could bring the  
status of the remaining viable MSP  solutions into sharper focus.   
These measurements would also provide important constraints on warm  
dark matter models or on the small scale power spectrum in CDM.

\section{Acknowledgments} 
We thank Jason Harris, Tobias Kaufmann, Savvas Koushiappas, Andrey Kravtsov, 
Steve Majewski, Nicolas Martin, Josh Simon, and Andrew Zentner for discussions on this topic. 
We thank Mike Siegal for sharing his Leo II data. 
LES is supported in part by a Gary McCue postdoctoral fellowship
through the Center for  Cosmology at the University of California,
Irvine. L.E.S., J.S.B., and M.K. are supported in part by NSF grant
AST-0607746. M.K. acknowledges support from PHY-0555689.    
J. D. acknowledges support from
NASA through Hubble Fellowship grant HST-HF-01194.01 awarded by the
Space Telescope Science Institute, which is operated by the
Association of Universities for Research in Astronomy, Inc., for NASA,
under contract NAS 5-26555.
P.M. acknowledges support from NASA grants
NAG5-11513 and NNG04GK85G, and from the Alexander von Humboldt
Foundation. The Via Lactea simulation was performed on NASA's Project 
Columbia supercomputer system.

\bibliography{ms}

\end{document}